\documentclass[amsfonts, amssymb, amsmath, aps, 10pt, pra, twocolumn, superscriptaddress, english, floatfix]{revtex4-2}
\usepackage[utf8]{inputenc}
\usepackage[english]{babel}
\usepackage[T1]{fontenc}
\usepackage{hyperref}
\usepackage[nolist]{acronym}
\usepackage{mathdots}
\usepackage{bm}
\usepackage{graphicx}
\usepackage{physics}
\usepackage{dsfont}
\usepackage{mathtools}
\usepackage[dvipsnames]{xcolor}

\hypersetup{
    colorlinks = true,
    linkcolor = RoyalBlue,
    citecolor = RoyalBlue,
    urlcolor = RoyalBlue
    }

\graphicspath{{figures/}}

\newcommand{\PTr}[2]{\text{Tr}_{#1}\left\{#2\right\}}

\newcommand{\cF}{\mathcal{F}}
\newcommand{\cZ}{\mathcal{Z}}
\newcommand{\cS}{\mathcal{S}}
\newcommand{\cO}{\mathcal{O}}
\newcommand{\cH}{\mathcal{H}}
\newcommand{\cP}{\mathcal{P}}

\newcommand{\dI}{\mathds{1}}

\begin{document}

\title{Variational Gibbs State Preparation on NISQ Devices}

\author{Mirko Consiglio}
\email{mirko.consiglio@um.edu.mt}
\affiliation{Department of Physics, University of Malta, Msida MSD 2080, Malta}

\author{Jacopo Settino}
\affiliation{Dipartimento di Fisica, Universit\`a della Calabria, 87036 Arcavacata di Rende (CS), Italy}
\affiliation{INFN, gruppo collegato di Cosenza, 87036 Arcavacata di Rende (CS), Italy}
\affiliation{ICAR-CNR, 87036 Rende (CS), Italy}

\author{Andrea Giordano}
\affiliation{ICAR-CNR, 87036 Rende (CS), Italy}

\author{Carlo Mastroianni}
\affiliation{ICAR-CNR, 87036 Rende (CS), Italy}

\author{Francesco Plastina}
\affiliation{Dipartimento di Fisica, Universit\`a della Calabria, 87036 Arcavacata di Rende (CS), Italy}
\affiliation{INFN, gruppo collegato di Cosenza, 87036 Arcavacata di Rende (CS), Italy}

\author{Salvatore Lorenzo}
\affiliation{Universit\`a degli Studi di Palermo, Dipartimento di Fisica e Chimica - Emilio Segr\`e, via Archirafi 36, I-90123 Palermo, Italy}

\author{Sabrina Maniscalco}
\affiliation{QTF Centre of Excellence, Department of Physics, University of Helsinki, P.O. Box 43, FI-00014 Helsinki, Finland}
\affiliation{Algorithmiq Ltd, Kanavakatu 3 C, FI-00160 Helsinki, Finland}

\author{John Goold}
\affiliation{School of Physics, Trinity College Dublin, College Green, Dublin 2, Ireland}

\author{Tony J. G. Apollaro}
\affiliation{Department of Physics, University of Malta, Msida MSD 2080, Malta}

\begin{abstract}
The preparation of an equilibrium thermal state of a quantum many-body system on \ac{NISQ} devices is an important task in order to extend the range of applications of quantum computation. Faithful Gibbs state preparation would pave the way to investigate protocols such as thermalization and out-of-equilibrium thermodynamics, as well as providing useful resources for quantum algorithms, where sampling from Gibbs states constitutes a key subroutine. We propose a \ac{VQA} to prepare Gibbs states of a quantum many-body system. The novelty of our \ac{VQA} consists in implementing a parameterized quantum circuit acting on two distinct, yet connected (via \textsc{CNOT} gates), quantum registers. The \ac{VQA} evaluates the Helmholtz free energy, where the von Neumann entropy is obtained via post-processing of computational basis measurements on one register, while the Gibbs state is prepared on the other register, via a unitary rotation in the energy basis. Finally, we benchmark our \ac{VQA} by preparing Gibbs states of the transverse field Ising and Heisenberg \textit{XXZ} models and achieve remarkably high fidelities across a broad range of temperatures in statevector simulations. We also assess the performance of the \ac{VQA} on IBM quantum computers, showcasing its feasibility on current \ac{NISQ} devices.
\end{abstract}

\maketitle

\begin{acronym}
\acro{PQC}{parameterized quantum circuit}
\acro{VQA}{variational quantum algorithm}
\acro{VQE}{variational quantum eigensolver}
\acro{NISQ}{noisy intermediate-scale quantum}
\acro{BFGS}{Broyden--Fletcher--Goldfarb--Shanno}
\acro{SPSA}{simultaneous perturbation stochastic approximation}
\acro{TFD}{thermofield double}
\acro{ML}{maximum likelihood}
\acro{NSB}{Nemenman--Shafee--Bialek}
\end{acronym}

\section{Introduction}
\acresetall

An integral task in quantum state preparation is the generation of finite-temperature thermal states of a given Hamiltonian, on a quantum computer. Indeed, Gibbs states (also known as thermal states) can be used for quantum simulation~\cite{Childs2018}, quantum machine learning~\cite{Kieferova2017, Biamonte2017}, quantum optimization~\cite{Somma2008}, and the study of open quantum systems~\cite{Poulin2009}. In particular, combinatorial optimization problems~\cite{Somma2008}, semi-definite programming~\cite{Brandao2016}, and training of quantum Boltzmann machines~\cite{Kieferova2017}, can be tackled by sampling from well-prepared Gibbs states.

The preparation of an arbitrary initial state is a challenging task in general, with finding the ground-state of a Hamiltonian being a QMA-hard problem~\cite{Watrous2008}. Preparing Gibbs states, specifically at low temperatures, could be as hard as finding the ground-state of that Hamiltonian~\cite{Aharonov2013}. The first algorithms for preparing Gibbs states were based on the idea of coupling the system to a register of ancillary qubits, and letting the system and environment evolve under a joint Hamiltonian, simulating the physical process of thermalization, such as in Refs.~\cite{Terhal2000, Poulin2009, Riera2012}, while others relied on dimension reduction~\cite{Bilgin2010}.

The algorithm proposed in this manuscript, for preparing Gibbs states, can be placed in the category of \acp{VQA}, such as in Refs.~\cite{Wu2019, Chowdhury2020, Wang2021,  Warren2022, Foldager2022}, and similarly for preparing \ac{TFD} states~\cite{Zhu2020, Premaratne2020, Sagastizabal2021}. Variational ans\"atze based on multi-scale entanglement renormalization~\cite{Sewell2022} and product spectrum ansatz~\cite{Martyn2019} have also been proposed in order to prepare Gibbs states.

Alternative algorithms prepare thermal states through quantum imaginary time evolution, such as in Refs.~\cite{Verstraete2004, Chowdhury2016, Zoufal2021, Gacon2021, Wang2023}, starting from a maximally mixed state, while others start from a maximally entangled state~\cite{Yuan2019}. Ref.~\cite{Haug2022} proposed quantum-assisted simulation to prepare thermal states, which does not require a hybrid quantum--classical feedback loop. In addition, methods exist to sample Gibbs state expectation values, rather than prepare the Gibbs state directly, such as in quantum metropolis methods~\cite{Temme2011, Yung2012}, imaginary time evolution applied to pure states~\cite{Motta2020}, and random quantum circuits using intermediate measurements~\cite{Shtanko2023}.

Recent methods also propose using rounding promises~\cite{Rall2023}, fluctuation theorems~\cite{Holmes2022}, pure thermal shadow tomography~\cite{Coopmans2023}, and minimally entangled typical thermal states for finite temperature simulations~\cite{Getelina2023}.

The goal of this work is to propose a \ac{VQA} that efficiently prepares Gibbs states on \ac{NISQ} computers, employing the free energy as a (physically-motivated) objective function. This requires the evaluation of the von Neumann entropy~\cite{Bengtsson2006}, which is generally hard to obtain from a quantum register. In contrast to some of the currently employed \acp{VQA}~\cite{Riera2012, Wu2019, Chowdhury2020, Zhu2020, Wang2021, Shtanko2023, Sagastizabal2021, Warren2022}, which employ truncated equations to approximate it, we directly estimate the von Neumann entropy, without any truncation and with an error solely dependent on the number of shots, using sufficiently expressible ans\"atze capable of preparing the Boltzmann distribution. Our \ac{VQA}, in fact, is composed of two ans\"atze: a heuristic, shallow one that prepares the Boltzmann distribution, for a given temperature; and another one, possibly designed with a problem-inspired approach, which depends on the Hamiltonian, while being independent of the temperature. 

The paper is organized as follows: in Section~\ref{sec:GSP}, we present the \ac{VQA} for preparing Gibbs states; in Section~\ref{sec:results_ising}, we apply the algorithm to the Ising model, using both statevector and noisy simulations, as well as running the algorithm on IBM quantum hardware; in Section~\ref{sec:results_heisenberg}, we apply the algorithm to the Heisenberg model, using both statevector and shot-based simulations; and finally, in Section~\ref{sec:conclusion}, we draw our conclusions and discuss future prospects of this work.

\section{Variational Gibbs State Preparation} \label{sec:GSP}

Consider a Hamiltonian $\cH$, describing $n$ interacting qubits, then, the Gibbs state at inverse temperature $\beta \equiv 1 / \left(k_\text{B} T\right)$, where $k_\text{B}$ is the Boltzmann constant and $T$ is the temperature, is defined as
\begin{equation}
    \rho(\beta, \cH) = \frac{e^{-\beta \cH}}{\cZ(\beta, \cH)},
\end{equation}
where the partition function $\cZ(\beta, \cH)$ is
\begin{equation}
    \cZ(\beta, \cH) = \Tr{e^{-\beta \cH}} = \sum_{i=0}^{d - 1} e^{-\beta E_i}.
\end{equation}
Here  $d = 2^n$, while $\{E_i\}$ are the eigenenergies of $\cH$, with $\{\ket{E_i}\}$ denoting the corresponding eigenstates, i.e. $\cH\ket{E_i} = E_i\ket{E_i}$. 

Fixing a Hamiltonian $\cH$ and inverse temperature $\beta$, for a general state $\rho$, one can define a generalized Helmholtz free energy as
\begin{equation}
    \cF(\rho) = \Tr{\cH \rho} - \beta^{-1}\cS(\rho),
    \label{eq:helmholtz_free_energy}
\end{equation}
where the von Neumann entropy $\cS(\rho)$ can be expressed in terms of the eigenvalues, $p_i$, of $\rho$,
\begin{equation}
    \cS(\rho) = -\sum_{i=0}^{d - 1} p_i \ln p_i.
    \label{eq:von_neumann}
\end{equation}
Since the Gibbs state is the unique state that minimizes the free energy of $\cH$~\cite{Matsui1994}, a variational procedure can be put forward that takes Eq.~\eqref{eq:helmholtz_free_energy} as an objective function, such that
\begin{equation}
    \rho(\beta, \cH) = \underset{\rho}{\arg\min}~\cF(\rho).
\end{equation}
In this case, $p_i = \exp\left(-\beta E_i\right)/\cZ(\beta, \cH)$ is the probability of getting the eigenstate $\ket{E_i}$ from the ensemble $\rho(\beta, \cH)$.

\subsection{Framework of the Algorithm}

The difficulty in measuring the von Neumann entropy, defined by Eq.~\eqref{eq:von_neumann}, of a quantum state on a \ac{NISQ} device is typically the challenging part of variational Gibbs state preparation algorithms, as $\cS(\rho)$ is not an observable. With this in mind, we present a \ac{VQA} that avoids the direct measurement of the von Neumann entropy on a quantum computer, by using a carefully constructed \ac{PQC}.

When preparing an $n$-qubit state starting from the input state $\ket{0}^{\otimes n}$, given that a quantum computer operates using only unitary gates, the final quantum state of the entire register will be pure. As a result, in order to prepare an $n$-qubit Gibbs state on the system register, we require an $m \leq n$-qubit ancillary register. For example, in the case of the infinite-temperature Gibbs state, which is the maximally mixed state, we require $m = n$ qubits in the ancillary register to achieve maximal von Neumann entropy. In order to evaluate the von Neumann entropy, without any truncation, we need to be able to prepare the entire Boltzmann distribution on the ancillary register, hence, we set $m = n$, irrespective of the temperature.

We shall denote the ancillary register as $A$, while the preparation of the Gibbs state will be carried out on the system register $S$. The purpose of the \ac{VQA} is to effectively create the Boltzmann distribution on $A$, which is then imposed on $S$, via intermediary \textsc{CNOT} gates, to generate a diagonal mixed state. In the ancillary register we can choose a unitary ansatz capable of preparing such a probability distribution. Thus, the ancillary qubits are responsible for mixing in the probabilities of the thermal state, while also being able to access these probabilities via measurements in the computational basis. On the other hand, the system register will host the preparation of the Gibbs state as well as the measurement of the expectation value of our desired Hamiltonian.

The specific design of the \ac{PQC} instead allows classical post-processing of simple measurement results, carried out on ancillary qubits in the computational basis, to determine the von Neumann entropy. A diagrammatic representation of the structure of the \ac{PQC} is shown in Fig.~\ref{fig:gibbs_circuit}. Note that while the \ac{PQC} of the algorithm has to have a particular structure --- a unitary acting on the ancillae and a unitary acting on the system, connected by intermediary \textsc{CNOT} gates --- it is not dependent on the choice of Hamiltonian $\cH$, inverse temperature $\beta$, or the variational ans\"atze, $U_A$ and $U_S$, employed within.

\subsection{Modular Structure of the PQC}

\begin{figure}[t]
    \centering
    \includegraphics[width=0.35\textwidth]{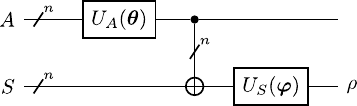}
    \caption{\ac{PQC} for Gibbs state preparation, with systems $A$ and $S$ each carrying $n$ qubits. \textsc{CNOT} gates act between each qubit $A_i$ and corresponding $S_i$.}
    \label{fig:gibbs_circuit}
\end{figure}

The \ac{PQC}, as shown in Fig.~\ref{fig:gibbs_circuit} for the \ac{VQA}, is composed of a unitary gate $U_A$ acting on the ancillary qubits, and a unitary gate $U_S$ acting on the system qubits, with \textsc{CNOT} gates in between. Note that the circuit notation we are using here means that there are $n$ qubits for both the system and the ancillae, as well as $n$ \textsc{CNOT} gates that act in parallel, and are denoted as
\begin{equation}
    \textsc{CNOT}_{AS} \equiv \bigotimes\limits_{i=0}^{n - 1}\textsc{CNOT}_{A_i S_i}.
\end{equation}

The parameterized unitary $U_A$ acting on the ancillae, followed by \textsc{CNOT} gates between the ancillary and system qubits, is responsible for preparing a probability distribution on the system. The parameterized unitary $U_S$ is then applied on the system qubits to transform the computational basis states into the eigenstates of the Hamiltonian.

A general unitary gate of dimension $d = 2^n$ is given by
\begin{equation}
    U_A = \left( \begin{array}{cccc}
        u_{0,0} & u_{0,1} & \cdots & u_{0,d-1} \\
        u_{1,0} & u_{1,1} & \cdots & u_{1,d-1} \\
        \vdots & \vdots & \ddots & \vdots \\
        u_{d-1,0} & u_{d-1,1} & \cdots & u_{d-1,d-1}
        \end{array} \right).
    \label{eq:UA}
\end{equation}
Starting with the initial state of the $2n$-qubit register, $\ket{0}_{AS}^{\otimes 2n}$, we apply the unitary gate $U_A$ on the ancillae to get a quantum state $\ket{\psi}_A$, such that
\begin{equation}
    (U_A \otimes \dI_S) \ket{0}_{AS}^{\otimes 2n} = \ket{\psi}_{A} \otimes \ket{0}_{S}^{\otimes n},
    \label{eq:U_A}
\end{equation}
where
\begin{equation}
    \ket{\psi}_A = \sum_{i=0}^{d - 1} u_{i,0} \ket{i}_A,
    \label{eq:psi_A}
\end{equation}
and $\dI_S$ is the identity operator acting on the system. The next step is to prepare a probability mixture on the system qubits, which can be done by applying \textsc{CNOT} gates between each ancilla and system qubit. This results in a state
\begin{align}
    \textsc{CNOT}_{AS} \left( \ket{\psi}_{A} \otimes \ket{0}_{S}^{\otimes n} \right) = \sum_{i=0}^{d - 1} u_{i,0} \ket{i}_A \otimes \ket{i}_S.
    \label{eq:pre_tfd}
\end{align}
By then tracing out the ancillary qubits, we arrive at
\begin{align}
    &\PTr{A}{\left( \sum_{i=0}^{d-1} u_{i,0}\ket{i}_A \otimes \ket{i}_S \right) \left( \sum_{j=0}^{d-1} u_{j,0}^* \bra{j}_A \otimes \bra{j}_S \right)} \nonumber \\ = \
    &\sum_{i,j=0}^{d-1} u_{i,0}u_{j,0}^* \braket{i}{j} \ketbra{i}{j}_S = \
    \sum_{i=0}^{d-1} |u_{i,0}|^2 \ketbra{i}{i}_S,
\end{align}
ending up with a diagonal mixed state on the system, with probabilities given directly by the absolute square of the entries of the first column of $U_A$, that is, $p_i = |u_{i,0}|^2$. If the system qubits were traced out instead, we would end up with the same diagonal mixed state, but on the ancillary qubit register:
\begin{align}
    &\PTr{S}{\left( \sum_{i=0}^{d-1} u_{i,0}\ket{i}_A \otimes \ket{i}_S \right) \left( \sum_{j=0}^{d-1} u_{j,0}^* \bra{j}_A \otimes \bra{j}_S \right)} \nonumber \\ = \
    &\sum_{i,j=0}^{d-1} u_{i,0}u_{j,0}^* \braket{i}{j} \ketbra{i}{j}_A = \
    \sum_{i=0}^{d-1} |u_{i,0}|^2 \ketbra{i}{i}_A,
\end{align}
This implies that by measuring in the computational basis of the ancillary qubits, we can determine the probabilities $p_i$, which can then be post-processed to determine the von Neumann entropy $\cS$ of the state $\rho$ via Eq.~\eqref{eq:von_neumann} (since the entropy of $A$ is the same as that of $S$). As a result, since $U_A$ only serves to create a probability distribution from the entries of the first column, we can do away with a parameterized orthogonal (real unitary) operator, thus requiring less gates and parameters for the ancillary ansatz.

The unitary gate $U_S$ then serves to transform the computational basis states of the system qubits to the eigenstates of the Gibbs state, such that
\begin{align}
    \rho &= U_S \left( \sum_{i=0}^{d-1} |u_{i,0}|^2 \ketbra{i}{i}_S \right) U_S^\dagger \nonumber \\ 
    &= \sum_{i=0}^{d - 1} p_i \ketbra{\psi_i},
\end{align}
where the expectation value $\Tr{\cH \rho}$ of the Hamiltonian can be measured. Ideally, at the end of the optimization procedure, $p_i = \exp\left(-\beta E_i\right)/\cZ(\beta, \cH)$ and $\ket{\psi_i} = \ket{E_i}$, so that we get
\begin{equation}
    \rho(\beta, \cH) = \sum_{i=0}^{d - 1} \frac{e^{-\beta E_i}}{\cZ(\beta, \cH)} \ketbra{E_i}.
\end{equation}
The \ac{VQA} therefore avoids the entire difficulty of measuring the von Neumann entropy of a mixed state on a quantum computer, and instead transfers the task of post-processing computational basis measurement results to the classical computer, which is much more tractable.

\subsection{Objective Function}

Finally, we can define the objective function of our \ac{VQA} to minimize the free energy~\eqref{eq:helmholtz_free_energy}, via our constructed \ac{PQC}, to obtain the Gibbs state
\begin{align}
    \rho(\beta, \cH) &= \underset{\bm{\theta}, \bm{\varphi}}{\arg\min} \ \cF\left(\rho\left(\bm{\theta}, \bm{\varphi}\right)\right) \nonumber \\ &= \underset{\bm{\theta}, \bm{\varphi}}{\arg\min} \left( \Tr{\cH \rho_S(\bm{\theta}, \bm{\varphi})} - \beta^{-1}\cS\left(\rho_A(\bm{\theta})\right) \right).
    \label{eq:gibbs_state}
\end{align}

It is noteworthy to mention that while the energy expectation depends on both sets of angles $\bm{\theta}$ (as $U_A$ is responsible for parameterizing the Boltzmann distribution) and $\bm{\varphi}$ (as $U_S$ is responsible for parameterizing the eigenstates of the Gibbs state), the calculation of the von Neumann entropy only depends on $\bm{\theta}$.

\begin{figure}[t]
    \centering
    \includegraphics[width=0.42\textwidth]{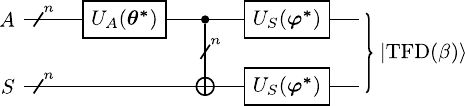}
    \caption{Optimal \ac{PQC} for \ac{TFD} state preparation, with systems $A$ and $S$ each carrying $n$ qubits. \textsc{CNOT} gates act between each qubit $A_i$ and corresponding $S_i$.}
    \label{fig:tfd_circuit}
\end{figure}

Furthermore, once we obtain the optimal parameters $\bm{\theta^*}$, $\bm{\varphi^*}$, preparing the Gibbs state $\rho(\beta, \cH)$ on the system qubits $S$, one can place the same unitary $U_S$ with optimal parameters $\bm{\varphi^*}$ on the ancillary qubits to prepare the \ac{TFD} state on the entire qubit register, as shown in Fig.~\ref{fig:tfd_circuit}. A \ac{TFD} state~\cite{Wu2019, Zhu2020, Sagastizabal2021} is defined as
\begin{equation}\label{eq:tfd}
    \ket{\textrm{TFD}(\beta)} = \sum_{i=0}^{d - 1} \sqrt{\frac{e^{-\beta E_i}}{\cZ(\beta, \cH)}} \ket{i}_A \otimes \ket{i}_S~,
\end{equation}
and, tracing out either the ancilla or system register, yields the same Gibbs state on the other register. Eq.~\eqref{eq:tfd} is equivalent to Eq.~\eqref{eq:pre_tfd} after applying $U_S$ on both registers.

\subsection{Alternative Implementations of the Algorithm}

\begin{figure}[t]
    \centering
    \includegraphics[width=0.35\textwidth]{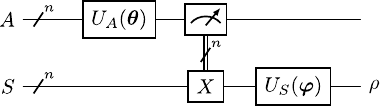}
    \caption{\ac{PQC} for Gibbs state preparation using mid-circuit measurements, with systems $A$ and $S$ each carrying $n$ qubits. Classically-controlled \textsc{NOT} gates act between each qubit $A_i$ and corresponding $S_i$.}
    \label{fig:classical_gibbs_state}
\end{figure}

There are several adjustments that could be applied to the \ac{PQC} to modify the procedure. One specific example is replacing the intermediary \textsc{CNOT} gates with mid-circuit measurements and implementing classically controlled-\textsc{NOT} gates, since no subsequent unitary gates act on the control qubits, as shown in Fig.~\ref{fig:classical_gibbs_state}. This method admits a few benefits:
\begin{enumerate}
    \item Since the ancillary system needs to be measured to compute the von Neumann entropy, utilizing mid-circuit measurements followed by classically-controlling the system qubits is a natural approach to the algorithm.
    \item The two registers $A$ and $S$ can be made fully distinct in terms of the device topology, as well as reducing the depth of the entire circuit, leading to less overall decoherence affecting the protocol.
    \item Once optimization is carried out, the classically-controlled \textsc{NOT} gates can still be kept in the circuit, yet if the experimentalist ignores the measurement information (equivalent to tracing out), then there is no operational difference between preparing an ensemble of pure states and preparing a mixed state using quantum \textsc{CNOT} gates.
\end{enumerate}
The only downside is if the ancillary qubits are intended to be used again, such as when preparing the \ac{TFD} state. In this case, the optimization for finding optimal parameters to prepare the Gibbs state can still be carried out using classically-controlled \textsc{NOT} gates. However, at the end of the optimization procedure, the classically-controlled \textsc{NOT} gates can be replaced with \textsc{CNOT} gates followed by the optimized system unitary, with the same structure as in Fig.~\ref{fig:tfd_circuit}, to obtain the \ac{TFD} state.

The \ac{VQA} can be further be adapted so that an $U_A$ is replaced by a classical procedure that generates the probability distribution, and prepares pure states of the Gibbs state ensemble $\{p_i, \ket{E_i}\}$, where $p_i = \exp(-\beta E_i) / \cZ$ on the system qubits $S$. This procedure can be carried out by parametrizing a classical probability distribution $p(\bm{\theta})$ by $\cO(\text{poly}(n))$ parameters $\bm{\theta}$, using methods such as Markov chains composed of a sequence of local stochastic matrices, among others. The probability distribution will output bit strings $\ket{i}$ that can be fed as a computational input state to the unitary $U_S$ that prepares the eigenstates $\ket{E_i}$ of the Hamiltonian. By reducing the number of qubits and eliminating the requirement for intermediary \textsc{CNOT} gates, this process may result in a less expressible probability distribution function because entanglement is not used as a resource.

Furthermore, if the parametrization of the probability distribution corresponds with the output distribution of a known unitary circuit, of a sufficiently shallow depth and expressibility, then the optimization can be carried using only the classical subroutine of sampling bit strings from the probability distribution $p(\bm{\theta})$, and feeding them to $U_S$ to compute the free energy. Once the \ac{VQA} is trained, $U_A$ can be introduced with the optimized parameters $\bm{\theta^*}$, to prepare the mixed Gibbs on the quantum computer. Nevertheless, finding such a parametrization, that corresponds to a shallow, yet expressible enough unitary, is a non-trivial task.

\section{Performance of the VQA on the Ising Model} \label{sec:results_ising}

In this Section we assess the performance of the \ac{VQA} for Gibbs state preparation of an Ising model. The Ising model is defined as
\begin{equation}
    \cH = -\sum_{i=1}^n \sigma_i^x \sigma_{i+1}^x - h \sum_{i=1}^n \sigma_i^z.
    \label{eq_Ising}
\end{equation}
The Ising Hamiltonian is a widely-investigated model~\cite{Franchini2016a}, and here we only report one relevant property for implementing a problem-inspired ansatz for $U_S$. The Hamiltonian in Eq.~\eqref{eq_Ising} commutes with the parity operator $\cP=\bigotimes_{i=0}^{n - 1} \sigma^z_i$. As a consequence, the eigenstates of $\cH$ have definite parity, and so will the eigenstates of $\rho_{\beta}$.

We use the Uhlmann-Josza fidelity~\cite{Uhlmann2011}, defined as $F\left(\rho, \sigma\right)=\left(\Tr{\sqrt{\sqrt{\rho}\sigma\sqrt{\rho}}}\right)^2$, as a figure of merit for the performance of our \ac{VQA}, since it describes how ``close'' the prepared state is to the Gibbs state, and it is also the most commonly-employed measure of distinguishability. However, other measures can be used, which have different interpretations. One example is the trace distance~\cite{Nielsen2010}, which enjoys the property that, if its value between the two states is bounded by $\epsilon$, expectation values computed on the effectively prepared state, differ from those taken on the Gibbs state by an amount that is, at most, proportional to $\epsilon$~\cite{Holmes2022}. Another choice is the relative entropy~\cite{Nielsen2010}, which describes the distinguishability between the two states as the surprise that occurs when an event happens that is not possible with the true Gibbs state~\cite{Vedral2002}.

We use a simple, linearly entangled \ac{PQC} for the unitary $U_A$, with parameterized $R_y(\theta_i)$ gates, and \textsc{CNOT}s as the entangling gates. This ansatz is hardware efficient and is sufficient to produce real amplitudes for preparing the probability distribution. Note that we require the use of entangling gates, as otherwise we will not be able to prepare any arbitrary probability distribution, including the Boltzmann distribution of the Ising model. A proof of this is given in Appendix~\ref{sec:ancilla_ansatz}.

\begin{figure*}[t]
    \centering
    \includegraphics[width=\textwidth]{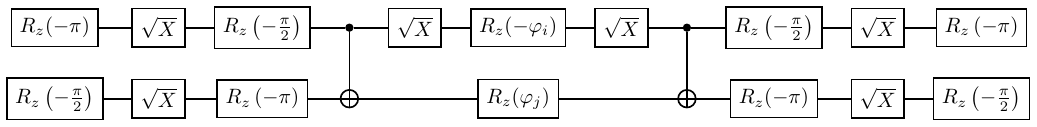}
    \caption{Decomposed $R_p$ gate in Eq.~\eqref{eq:R_p}.}
    \label{fig:R_p}
\end{figure*}

\begin{figure*}[t]
    \centering
    \includegraphics[width=\textwidth]{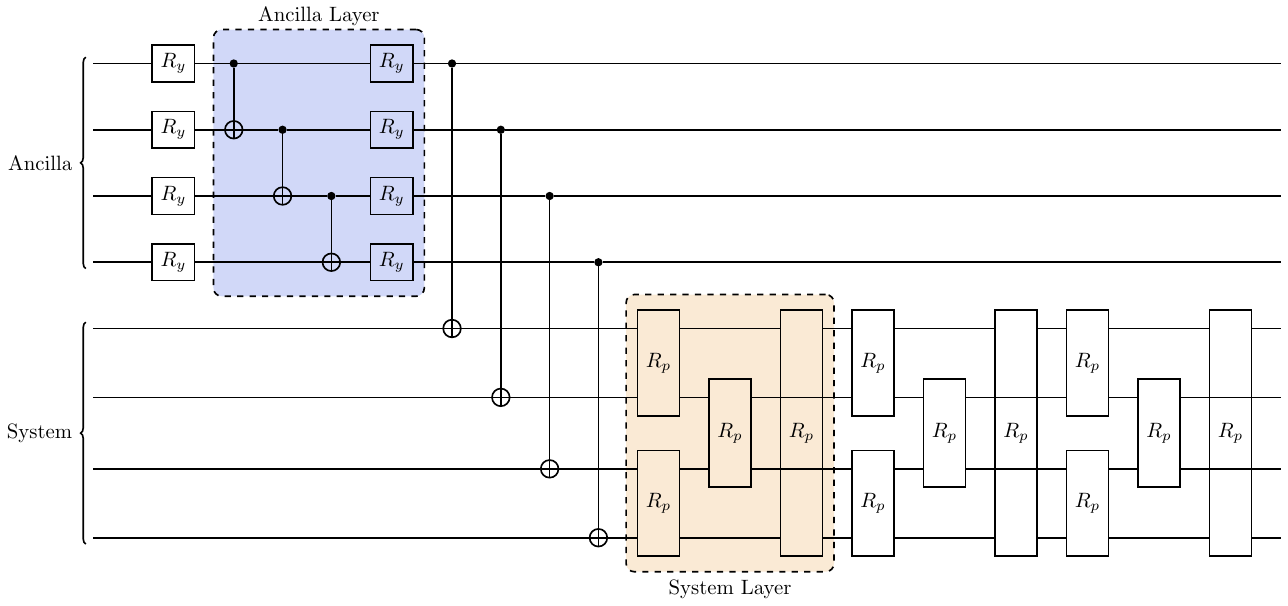}
    \caption{Example of an eight-qubit \ac{PQC}, consisting of one ancilla layer acting on a four-qubit register, and three ($n - 1$) system layers acting on another four-qubit register. Each $R_y$ gate is parameterized with one parameter $\theta_i$, while each $R_p$ gate has two parameters $\varphi_i$ and $\varphi_j$. The $R_p$ gate is defined in Eq.~\eqref{eq:R_p}. Note that the intermediary \textsc{CNOT} gates, as well as the $R_p$ gates acting on qubits two and three, and on qubits one and four of the system, can be carried out in parallel, respectively.}
    \label{fig:pqc_4qubits}
\end{figure*}

For the unitary $U_S$, we choose a parity-preserving \ac{PQC}. We employ a brick-wall structure, with the gates being $R_{xy}(\varphi_i) \equiv \exp(-\imath \varphi_i (\sigma^x \otimes \sigma^y)/2)$ followed by $R_{yx}(\varphi_j) \equiv \exp(-\imath \varphi_j (\sigma^y \otimes \sigma^x)/2)$ gates. If we combine the two gates, which we denote as $R_p(\varphi_i, \varphi_j)$, we get
\begin{widetext}
\begin{equation}
    R_p(\varphi_i, \varphi_j) = R_{yx}(\varphi_j)\cdot R_{xy}(\varphi_i) =
    \left(
    \begin{array}{cccc}
     \cos \left(\frac{\varphi_i +\varphi_j }{2}\right) & 0 & 0 & \sin \left(\frac{\varphi_i +\varphi_j }{2}\right) \\
     0 & \cos \left(\frac{\varphi_i -\varphi_j }{2}\right) & -\sin \left(\frac{\varphi_i -\varphi_j }{2}\right) & 0 \\
     0 & \sin \left(\frac{\varphi_i -\varphi_j }{2}\right) & \cos \left(\frac{\varphi_i -\varphi_j }{2}\right) & 0 \\
     -\sin \left(\frac{\varphi_i +\varphi_j }{2}\right) & 0 & 0 & \cos \left(\frac{\varphi_i +\varphi_j }{2}\right) \\
    \end{array}
    \right),
    \label{eq:R_p}
\end{equation}
\end{widetext}
which can be decomposed into two \textsc{CNOT} gates, six $\sqrt{X}$ gates and ten $R_z$ gates. One layer of the unitary acting on the system qubits consists of a brick-wall structure, composed of an even-odd sublayer of $R_p$ gates, followed by an odd-even sublayer of $R_p$ gates. The decomposed unitary is shown in Fig.~\ref{fig:R_p}. An example of a \ac{PQC}, split into a four-qubit ancillary register, and a four-qubit system register, is shown in Fig.~\ref{fig:pqc_4qubits}. Table~\ref{tab:scaling} shows the scaling of the \ac{VQA} assuming a closed ladder connectivity, for $n > 2$.

\begin{table}[t]
\centering
\resizebox{\linewidth}{!}{
\begin{tabular}{|l|l|l|} 
    \hline
    \# of parameters & $n(l_A + 1) + 2nl_S$ & $\cO(n(l_A + l_S))$ \\ 
    \hline
    \# of \textsc{CNOT} gates & $(n - 1)l_A + 2nl_S + n$ & $\cO(n(l_A + l_S))$ \\
    \hline
    \# of $\sqrt{X}$ gates & $2n(l_A + 1) + 6nl_S$ & $\cO(n(l_A + l_S))$ \\ 
    \hline
    Circuit depth & $(n + 1)l_A + Pl_S + 3$ & $\cO(nl_A + l_S)$ \\ 
    \hline
\end{tabular}
}
\caption{Scaling of the \ac{VQA} assuming a closed ladder connectivity, for $n > 2$ of the Ising model, where $l_A$ and $l_S$ are the number of ancilla ansatz and system ansatz layers, respectively, and $P$ is 12 when $n$ is even and 18 when $n$ is odd.}
\label{tab:scaling}
\end{table}

\subsection{Statevector Results} \label{sec:statevector_results}

\begin{figure*}[t]
    \centering
    \includegraphics[width=\textwidth]{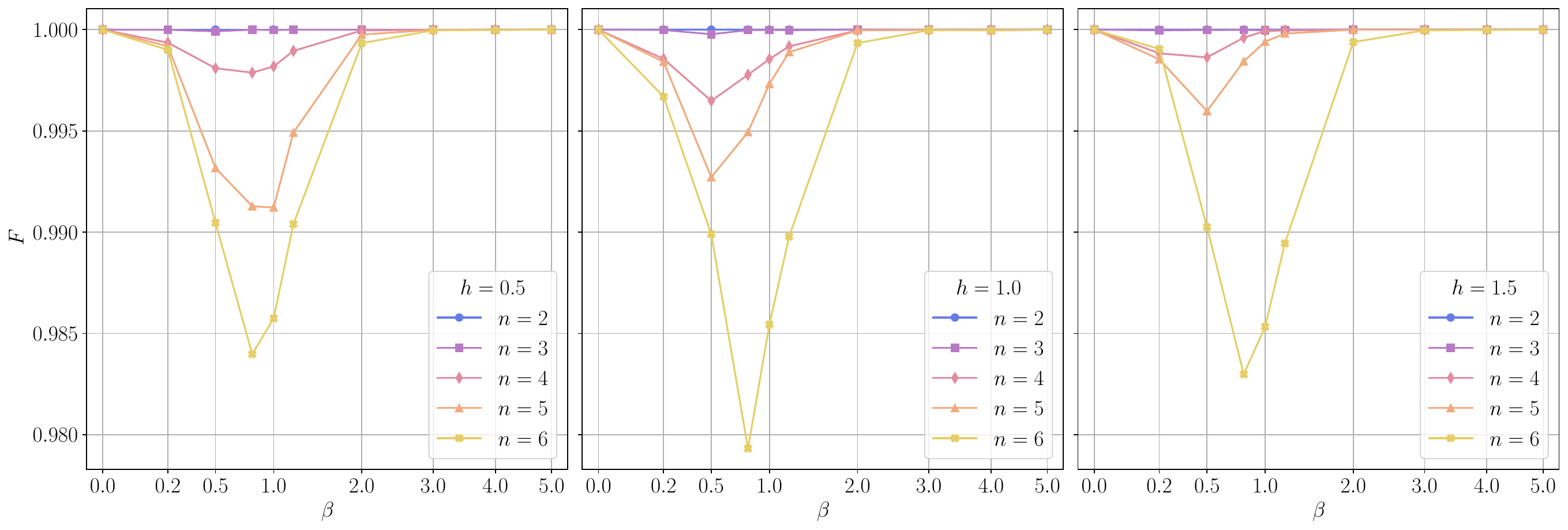}
    \caption{Fidelity $F$, of the obtained state via statevector simulations (using \ac{BFGS}) with the exact Gibbs state, vs inverse temperature $\beta$, for two to six qubits of the Ising model with $h = 0.5, 1.0, 1.5$. A total of 100 runs are made for each point, with the optimal state taken to be the one that maximizes the fidelity.}
    \label{fig:statevector_fidelity}
\end{figure*}

Fig.~\ref{fig:statevector_fidelity} shows the fidelity of the generated mixed state when compared with the exact Gibbs state of the Ising model with $h = 0.5, 1.0, 1.5$, respectively, across a range of temperatures for system size between two to six qubits. The \ac{VQA} was carried out using statevector simulations with the \ac{BFGS} optimizer~\cite{Nocedal2006}. We used one layer for the ancilla ansatz, and $n - 1$ layers for the system ansatz, with the scaling highlighted in Table~\ref{tab:scaling_2}. The number of layers was heuristically chosen to satisfy, at most, a polynomial scaling in quantum resources, while achieving a fidelity higher than 98\% in statevector simulations. Furthermore, in order to alleviate the issue of getting stuck in local minima, the optimizer is embedded in a Monte Carlo framework, that is, taking multiple random initial positions and carrying out a local optimization from each position --- which we call a `run' --- and finally taking the global minimum to be the minimum over all runs. 

\begin{table}[t]
\centering
\begin{tabular}{|l|l|l|} 
    \hline
    \# of parameters & $2n^2$ & $\cO(n^2)$ \\ 
    \hline
    \# of \textsc{CNOT} gates & $2n^2 - 1$ & $\cO(n^2)$ \\
    \hline
    \# of $\sqrt{X}$ gates & $2n(3n - 2)$ & $\cO(n^2)$ \\ 
    \hline
    Circuit depth & $(P + 1)n - P + 4$ & $\cO(n)$ \\
    \hline
\end{tabular}
\caption{Scaling of the \ac{VQA} assuming a closed ladder connectivity, for $n > 2$, with $l_A = 1$, and $l_S = n - 1$, and $P$ is 12 when $n$ is even and 18 when $n$ is odd. The depth counts both \textsc{CNOT} and $\sqrt{X}$ gates.}
\label{tab:scaling_2}
\end{table}

A total of 100 runs of \ac{BFGS} per $\beta$ were carried out to verify the reachability of the \ac{PQC}, with Fig.~\ref{fig:statevector_fidelity} showcasing the maximal fidelity achieved for each $\beta$ out of all runs. The results show that, indeed, our \ac{VQA}, is able to reach a very high fidelity $F \gtrsim 0.98$ for up to six-qubit Gibbs states of the Ising model. In the case of the extremal points, that is $\beta \rightarrow 0$ and $\beta \rightarrow \infty$, the fidelity reaches unity, for all investigated system sizes. However, for intermediary temperatures $\beta \sim 1$, the fidelity decreases with the number of qubits, which could be attributed to one layer of $U_A$ not being expressible enough to prepare the Boltzmann distribution around intermediary temperatures (since the von Neumann entropy depends solely on $U_A(\bm{\theta})$ as in Eq.~\eqref{eq:gibbs_state}). Moreover, at intermediary temperatures, most of the Boltzmann probabilities $p_i$ are still relatively small, resulting in a larger error in obtaining the correct eigenstate. On the other hand, at high temperatures, all probabilities are equally likely, and preparing the maximally mixed state is a straightforward task. While at low temperatures, the VQA effectively reduces to a \ac{VQE}, i.e., finding the ground-state of the Hamiltonian.

\subsection{Noisy Simulation Results} \label{sec:noisy_results}

The next step was to carry out noisy simulations of the \ac{VQA}. We took the noise model of \texttt{ibmq\_guadalupe}~\cite{IBM_compute_resources} for the Ising model with $h = 0.5$, similarly with one layer for the ancilla ansatz and $n - 1$ layers for the system ansatz. However, it must be noted that in this case, the scaling of the algorithm does not follow Table~\ref{tab:scaling_2}, due to the fact that \texttt{ibmq\_guadalupe} does not have a closed ladder connectivity. As a result, transpilation was carried out by the \texttt{Qiskit} transpiler using the SABRE algorithm~\cite{Li2018}. Apart from this, due to the \ac{BFGS} optimizer being incapable of optimizing a noisy objective function, an optimizer that accommodates noisy measurements was chosen: \ac{SPSA}~\cite{Spall1992}. 

\begin{figure}[t]
    \centering
    \includegraphics[width=0.48\textwidth]{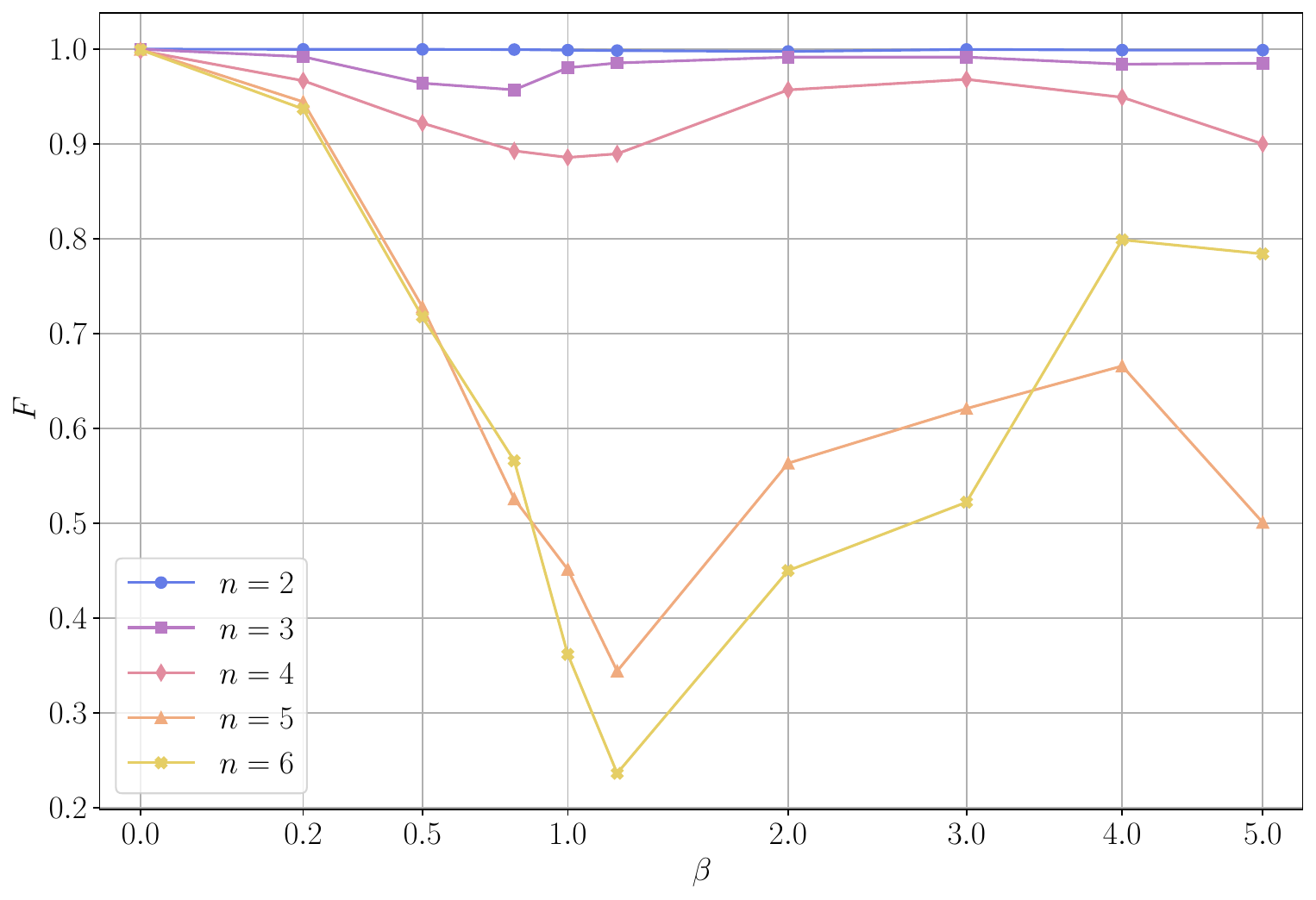}
    \caption{Fidelity $F$, of the obtained state via noisy simulations (using \ac{SPSA}) of \texttt{ibmq\_guadalupe} with the exact Gibbs state, vs inverse temperature $\beta$, for two to six qubits of the Ising model with $h = 0.5$. A total of ten runs are made for each point, with the optimal state taken to be the one that maximizes the fidelity.}
    \label{fig:noisy_fidelity}
\end{figure}

Using \ac{SPSA}, ten runs were carried out for each $\beta$, while the number of iterations was taken to be $100n$ for each run, with only two measurements at each iteration to estimate the gradient in a random direction, i.e. $200n$. As a consequence, a total of $2000n$ function evaluations were used to obtain the fidelity for each $\beta$ shown in Fig.~\ref{fig:noisy_fidelity} (with an extra 50 function evaluations at each run to calibrate the hyperparameters of \ac{SPSA}). Similar to the number of layers, the choice of the number of iterations was heuristically chosen so that, at most, the scaling is linear, while still retaining a fidelity greater than 95\% for the two- and three-qubit noisy simulation cases.

To measure the energy expectation value of the Ising model $\Tr{\cH\rho}$, we need to split the Ising Hamiltonian into its constituent Pauli strings, whose number scales linearly with the number of qubits as $2n$. However, we can group the $\sigma^x\sigma^x$ terms, as well as the $\sigma^z$ terms, and measure them simultaneously, reducing the number of measurement circuits to two. Each circuit was also measured with 1024 shots. The \texttt{M3} package~\cite{Nation2021} was also utilized to perform basic error mitigation. It operates by using a matrix-free preconditioned iterative-solution method to mitigate measurement error that does not form the full assignment matrix, or its inverse. A summary of the optimization scaling is shown in Table~\ref{tab:scaling_3}.

\begin{table}[t]
\centering
\resizebox{\linewidth}{!}{
\begin{tabular}{|l|l|l|}
    \hline
    \# of iterations for each run & $100n$ & $\cO(n)$ \\
    \hline
    \# of function evaluations for each run & $200n$ & $\cO(n)$ \\
    \hline
    \# of circuits per function evaluation & $2$ & $\cO(1)$ \\
    \hline
    \# of circuit evaluations for each run & $400n$ & $\cO(n)$ \\
    \hline
    \# of shots for each circuit evaluation & $1024$ & $\cO(1)$ \\
    \hline
\end{tabular}
}
\caption{Scaling of \ac{SPSA} for noisy simulations and on quantum hardware.}
\label{tab:scaling_3}
\end{table}

From Fig.~\ref{fig:noisy_fidelity}, one can see that the fidelity is significantly high in the case of $n = 2, 3, 4$. However, in the case of $n = 5, 6$, the Gibbs state is only faithfully prepared for low $\beta$. This can be attributed to the level of noise present in the optimization procedure, with the transpiled circuits going well beyond the quantum volume of \texttt{ibm\_guadalupe}. There are also some points which obtain a fidelity worse for five qubits than six qubits, which could be due to the larger depth acquired by an odd number of qubits in the system ansatz, as highlighted in Table~\ref{tab:scaling_2}.

Performance of \acp{VQA} is heavily impacted by the presence of noise-induced barren plateaus~\cite{Wang2021noise}. While the analysis of barren plateaus for Gibbs state preparation is beyond the scope of this manuscript, which aims at providing an alternative approach at variationally preparing Gibbs states, and avoiding any estimations of the entropy using Taylor expansions or other truncations, we still carry out brief analyses as starting points for future works. In particular, we discuss the implications of barren plateaus in Appendix~\ref{sec:barren_plateau}, and we also carry out analysis on the error of estimating the entropy in Appendix~\ref{sec:entropy_estimation}. 

\subsection{IBM Quantum Device Results} \label{sec:device_results}

\begin{figure}[t]
    \centering
    \includegraphics[width=0.48\textwidth]{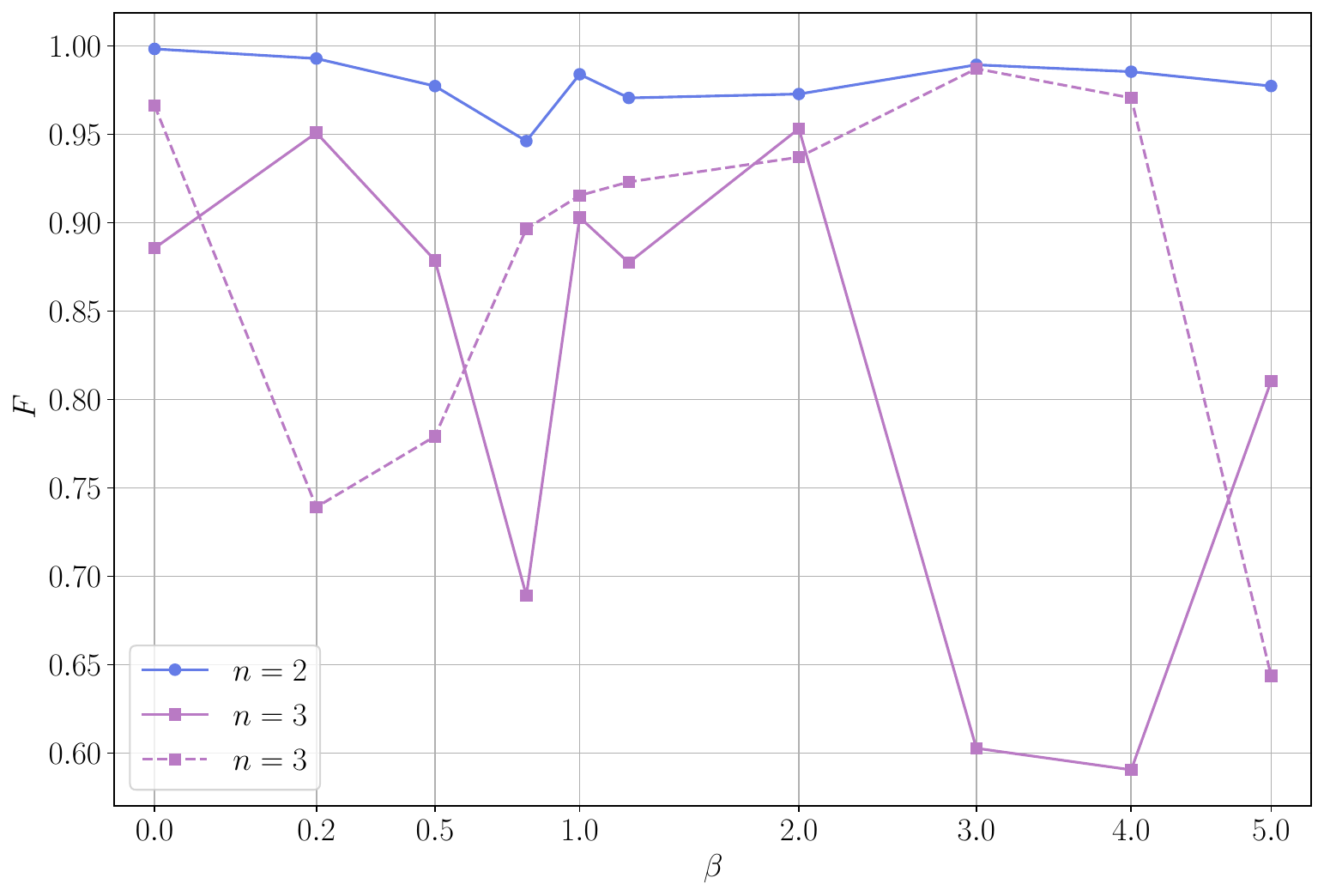}
    \caption{Fidelity $F$, of the obtained state (using \ac{SPSA}) running directly on \texttt{ibm\_nairobi} with the exact Gibbs state, vs inverse temperature $\beta$, for two and three qubits of the Ising model with $h = 0.5$. The dashed line represents the run with no $R_p$ gate between non-adjacent qubits in the system layers. One run is carried out for $n = 2$, and $n = 3$ for the dashed line, and two runs for $n = 3$ for the solid line.}
    \label{fig:ibm_nairobi}
\end{figure}

Finally, the \ac{VQA} was carried out on an actual quantum device. Fig.~\ref{fig:ibm_nairobi} displays the fidelity of Gibbs states obtained by running on IBM quantum hardware~\cite{IBM_compute_resources}, specifically \texttt{ibm\_nairobi}. Similarly to the noisy simulations, \ac{SPSA} was used; however, this time, with only one run for each $\beta$ in the case $n = 2$, and two runs in the case $n = 3$, with $100n$ iterations and 1024 shots. The Gibbs states were obtained by taking the optimal parameters from the optimization carried out on \texttt{ibm\_nairobi}, and determining the statevector on a classical computer.

The solid lines in Fig.~\ref{fig:ibm_nairobi} represent the two- and three-qubit results. At all points, the two-qubit Gibbs state shows excellent fidelity. On the other hand, the three-qubit Gibbs state is remarkably reproduced at certain temperatures, while it is lacking at other points. Since \texttt{ibm\_nairobi} does not have a closed ladder connectivity, several \textsc{SWAP} gates are necessary for carrying out transpilation. In an attempt to reduce the number of \textsc{SWAP} gates, we carried out another run at each $\beta$, where we removed the $R_p$ gate acting on non-adjacent qubits in the system layers, with the result shown in the dashed line of Fig.~\ref{fig:ibm_nairobi} (note that this also resulted in less parameters and depth of the \ac{PQC}). A considerable improvement in fidelity is achieved at the points were fidelity was lacking in the previous case. Since the available running time on the quantum device was limited, the amount of runs is still too low to determine the reason as to why the Gibbs state was not achieved with a higher fidelity. Nevertheless, comparing the results of Fig.~\ref{fig:ibm_nairobi}, with the statevector results in Fig.~\ref{fig:statevector_fidelity} and with the noise-simulated results in Fig.~\ref{fig:noisy_fidelity}, we conclude that limited connectivity, combined with device noise, is severely hampering the effectiveness of the \ac{VQA}.

\begin{figure}[t]
    \centering
    \includegraphics[width=0.48\textwidth]{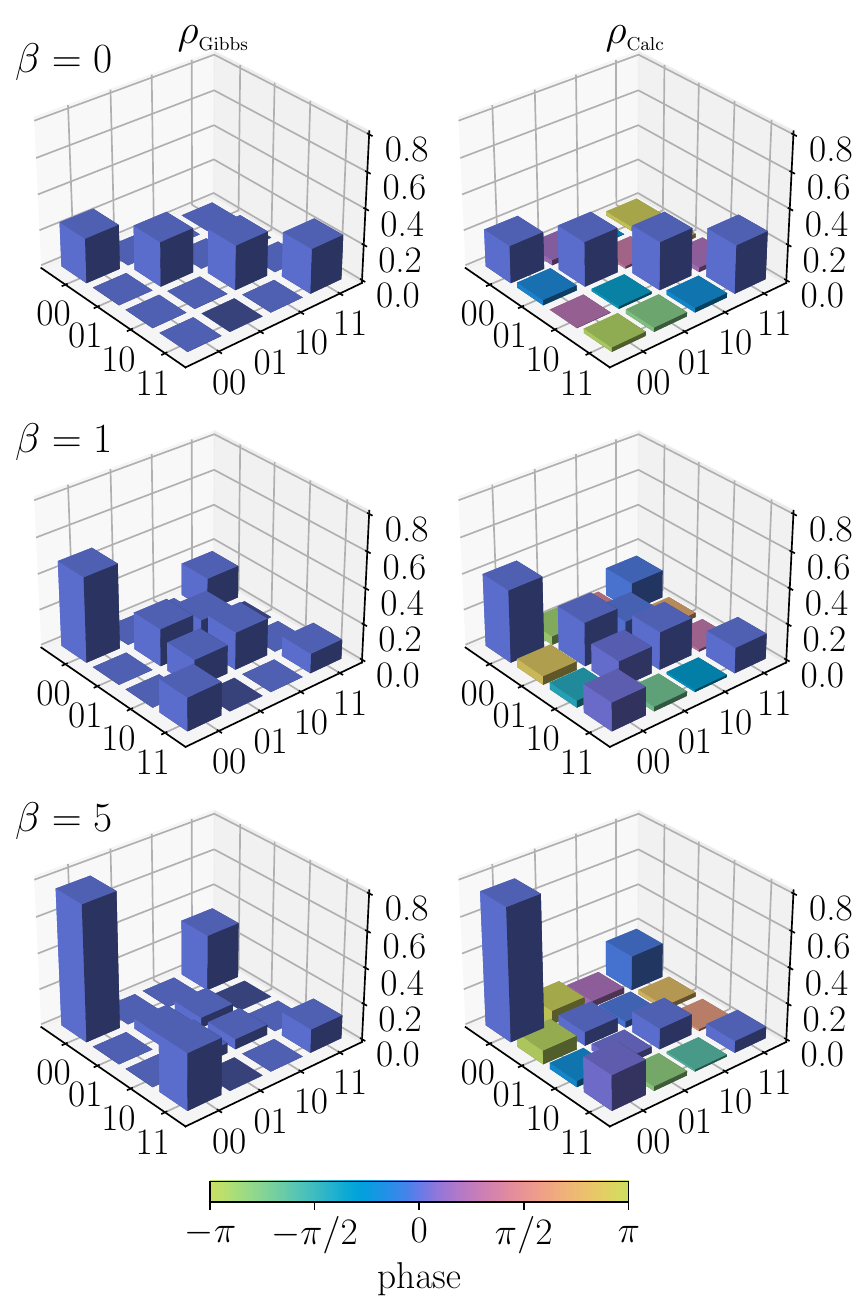}
    \caption{3D bar plot of the two qubit results from \texttt{ibm\_nairobi} for $\beta = 0, 1, 5$, of the Ising model with $h = 0.5$. The analytical Gibbs states are shown in the left column, while the tomographically obtained Gibbs States are shown in the right column.}
    \label{fig:bar3d}
\end{figure}

In addition, quantum state tomography for the two-qubit case was carried out on \texttt{ibm\_nairobi}, with 1024 shots, for the cases of $\beta = 0, 1, 5$, where the fidelities obtained were 0.992, 0.979, and 0.907, respectively. A 3D bar plot of the tomographic results can be seen in the right column Fig.~\ref{fig:bar3d}, and compared with the analytical form of the Gibbs state on the left column. The largest discrepancies can be witnessed in the off-diagonal terms, which increase as $\beta$ increases, as well as showcasing symptoms of amplitude damping. This could be attributed to the thermal relaxation and dephasing noise present in the quantum devices, leading to an overall decoherence in the Gibbs state.

While noisy simulations were ran using the calibration data of \texttt{ibmq\_guadalupe} --- since it has access of up to 16 qubits --- the actual hardware used for the two- and three-qubit Gibbs state preparation was \texttt{ibm\_nairobi}, due to its accessibility.

\section{Performance of the VQA on the Heisenberg \textit{XXZ} Model} \label{sec:results_heisenberg}

In this Section, to further explore the feasibility of our \ac{VQA} for few-body thermal state preparation on NISQ devices,  we assess its performance on a more complex, interacting system: the Heisenberg model. The Heisenberg \textit{XXZ} model is defined as
\begin{equation}
    \cH = -\frac{1}{4}\sum_{i=1}^n \left( \sigma_i^x \sigma_{i+1}^x + \sigma_i^y \sigma_{i+1}^y + \Delta \sigma_i^z \sigma_{i+1}^z \right) - h\sum_{i=1}^n \sigma_i^z.
    \label{eq_Heisenberg}
\end{equation}
At variance with the Ising model investigated in Section~\ref{sec:results_ising}, the so-called \textit{XXZ} model in a transverse magnetic field is an interacting Hamiltonian once mapped into spinless fermions. The phase diagram is much more complex and exhibits a paramagnetic-to-ferromagnetic phase transition at $h=\frac{1}{2}\left(1-\Delta\right)$~\cite{Franchini2016a}.

\begin{figure*}[t]
    \centering
    \includegraphics[width=\textwidth]{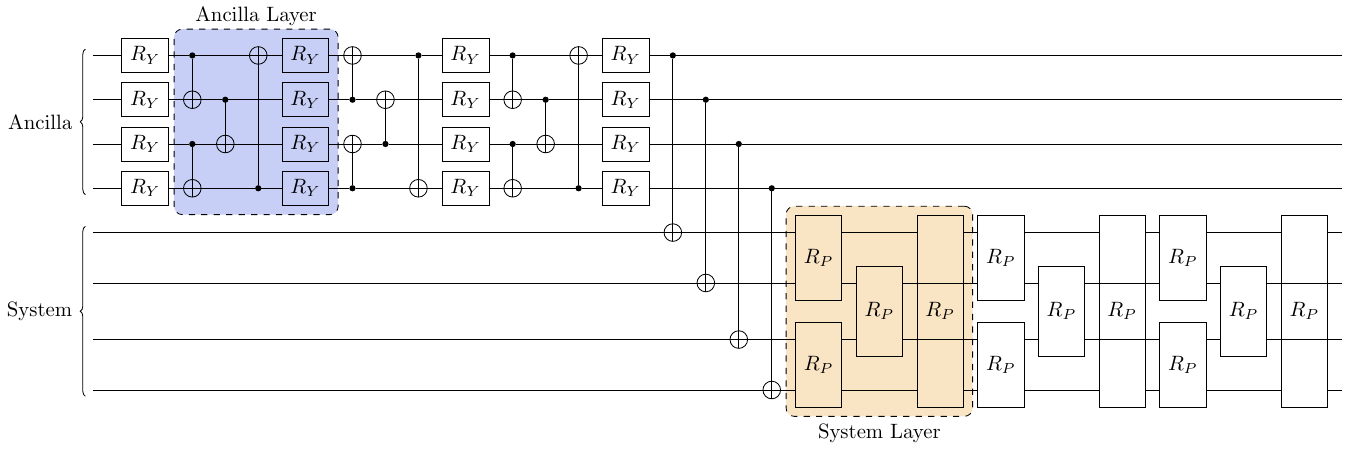}
    \caption{Example of an eight-qubit \ac{PQC}, consisting of three ($n - 1$) ancilla layers acting on a four-qubit register, and three ($n - 1$) system layers acting on another four-qubit register. Each $R_y$ gate is parameterized with one parameter $\theta_i$, while each $R_p$ gate has two parameters $\varphi_i$ and $\varphi_j$. The $R_p$ gate is defined in Eq.~\eqref{eq:R_p}. Note that the intermediary \textsc{CNOT} gates, as well as the $R_p$ gates acting on qubits two and three, and on qubits one and four of the system, can be carried out in parallel, respectively.}
    \label{fig:pqc_4qubits_heisenberg}
\end{figure*}

The Heisenberg model also commutes with the parity operator. As such, $U_S$ is the same as in section~\ref{sec:results_ising}. On the other hand, we use an alternating-layered ansatz for $U_A$, with parameterized $R_y(\theta_i)$ gates, and \textsc{CNOT}s as the entangling gates. Once again, this ansatz is hardware efficient and is sufficient to produce real amplitudes for preparing the probability distribution. We utilize the fidelity as our figure of merit for quantifying the performance of the algorithm. In this case, the scaling of $U_A$ and $U_S$ are given in Table~\ref{tab:scaling_4}. Moreover, an example of a \ac{PQC}, split into a four-qubit ancilla register, and a four-qubit system register, is shown in Fig.~\ref{fig:pqc_4qubits_heisenberg}.

\begin{table}[t]
\centering
\resizebox{\linewidth}{!}{
\begin{tabular}{|l|l|l|} 
    \hline
    \# of parameters & $n(l_A + 1) + 2nl_S$ & $\cO(n(l_A + l_S))$ \\ 
    \hline
    \# of \textsc{CNOT} gates & $nl_A + 2nl_S + n$ & $\cO(n(l_A + l_S))$ \\
    \hline
    \# of $\sqrt{X}$ gates & $2n(l_A + 1) + 6nl_S$ & $\cO(n(l_A + l_S))$ \\ 
    \hline
    Circuit depth & $Pl_A + 2Pl_S + 1$ & $\cO(l_A + l_S)$ \\ 
    \hline
\end{tabular}
}
\caption{Scaling of the \ac{VQA} assuming a closed ladder connectivity, for $n > 2$ of the Heisenberg model, where $l_A$ and $l_S$ are the number of ancilla ansatz and system ansatz layers, respectively, and $P$ is 2 when $n$ is even and 3 when $n$ is odd.}
\label{tab:scaling_4}
\end{table}

\subsection{Statevector Results}

\begin{figure*}[t]
    \centering
    \includegraphics[width=\textwidth]{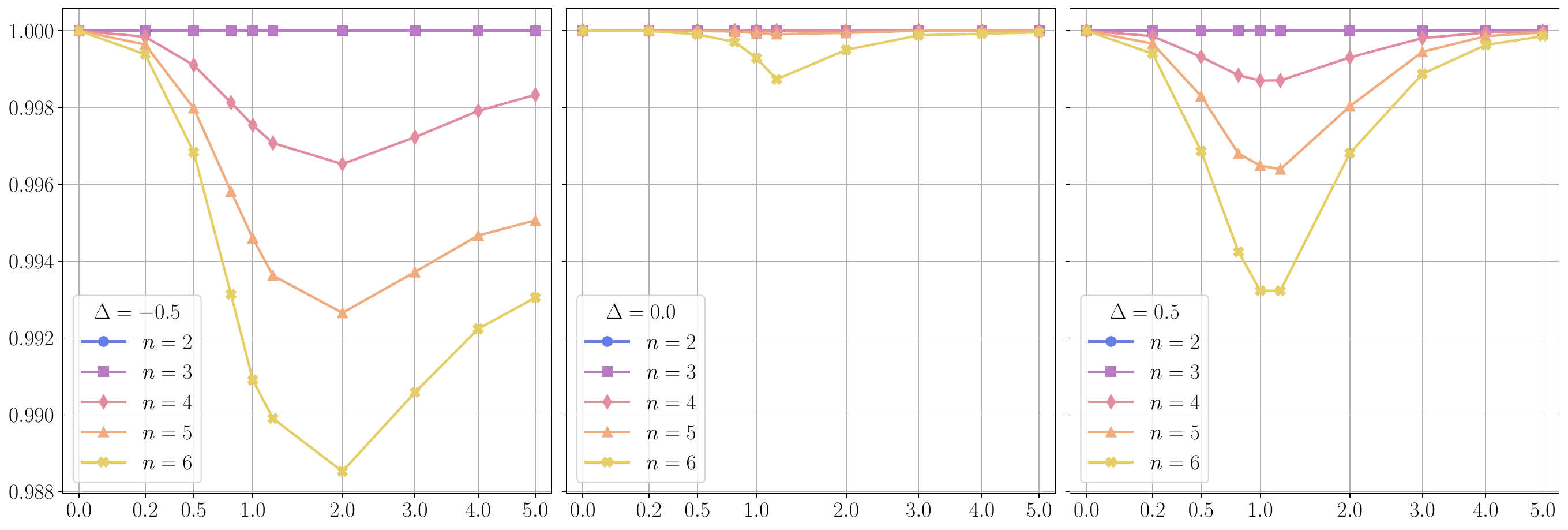}
    \caption{Fidelity $F$, of the obtained state via statevector simulations (using \ac{BFGS}) with the exact Gibbs state, vs inverse temperature $\beta$, for two to six qubits of the Heisenberg model with $h = 0.5$ and $\Delta = -0.5, 0.0, 0.5$. A total of 100 runs are made for each point, with the optimal state taken to be the one that maximizes the fidelity.}
    \label{fig:heisenberg_statevector}
\end{figure*}

Fig.~\ref{fig:heisenberg_statevector} shows the fidelity of the generated mixed state when compared with the exact Gibbs state of the Heisenberg model with $h = 0.5$ and $\Delta = -0.5, 0.0, 0.5$, respectively, across a range of temperatures for system size between two to six qubits. The \ac{VQA} was carried out using statevector simulations with the \ac{BFGS} optimizer~\cite{Nocedal2006}. We used $n - 1$ layers for the ancilla ansatz, and $n - 1$ layers for the system ansatz, with the scaling highlighted in Table~\ref{tab:scaling_5}. The number of layers was heuristically chosen to satisfy, at most, a polynomial scaling in quantum resources, while achieving a fidelity higher than 98\% in statevector simulations. Furthermore, in order to alleviate the issue of getting stuck in local minima, the optimizer is embedded in a Monte Carlo framework.

\begin{table}[t]
\centering
\begin{tabular}{|l|l|l|} 
    \hline
    \# of parameters & $3n^2 - 2n$ & $\cO(n^2)$ \\ 
    \hline
    \# of \textsc{CNOT} gates & $3n^2 - 2n$ & $\cO(n^2)$ \\
    \hline
    \# of $\sqrt{X}$ gates & $8n^2 - 6n$ & $\cO(n^2)$ \\ 
    \hline
    Circuit depth & $3Pn - 3P + 1$ & $\cO(n)$ \\
    \hline
\end{tabular}
\caption{Scaling of the \ac{VQA} assuming a closed ladder connectivity, for $n > 2$ for the Heisenberg model, with $l_A = n - 1$, and $l_S = n - 1$, and $P$ is 2 when $n$ is even and 3 when $n$ is odd. The depth counts only \textsc{CNOT} gates.}
\label{tab:scaling_5}
\end{table}

Similar to the statevector results of the Ising model in Section~\ref{sec:statevector_results}, we obtained very high fidelities $F > 0.98$ for a number of qubits ranging from two to six, across a broad range of temperatures of the Heisenberg \textit{XXZ} model. It must be noted that the paramagnetic-to-ferromagnetic transition point lies at $\Delta = 0$ for $h = 0.5$, resulting in the non-interacting $XX$ model, achieving a much better performance. The same dip in fidelity at intermediary temperatures reappears at around $\beta \sim 1$ for all the plots in Fig.~\ref{fig:heisenberg_statevector}.

\subsection{Shot-Based Results}

\begin{figure*}[t]
    \centering
    \includegraphics[width=\textwidth]{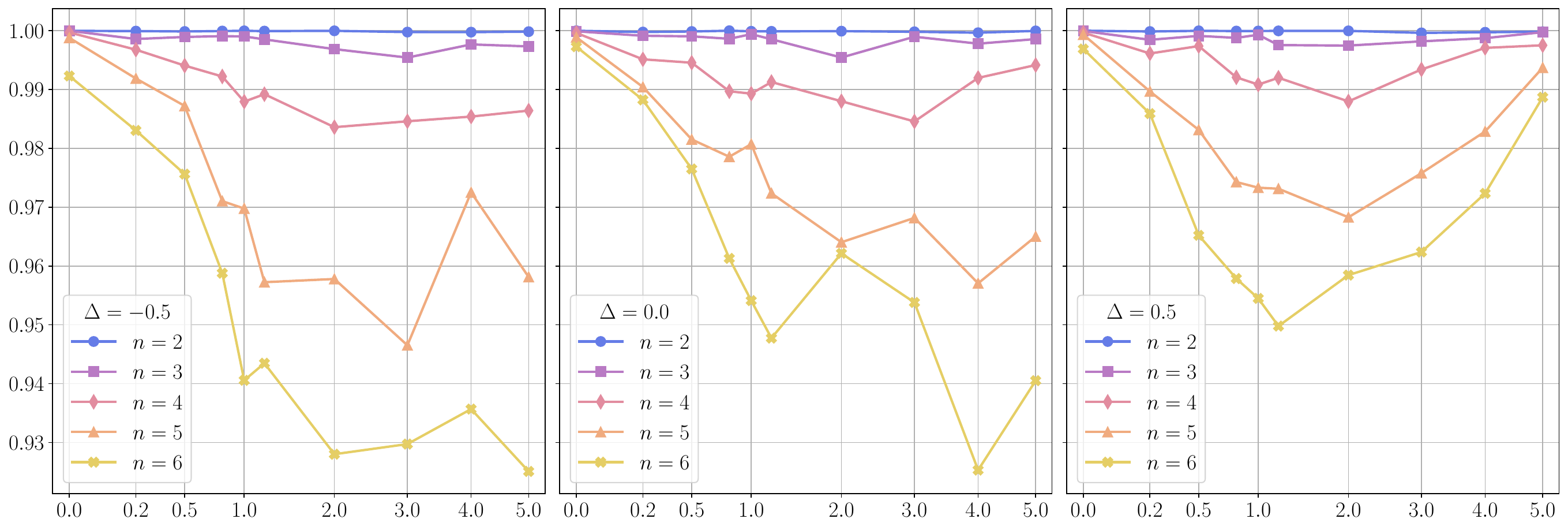}
    \caption{Fidelity $F$, of the obtained state via shot-based simulations (using \ac{SPSA}) with the exact Gibbs state, vs inverse temperature $\beta$, for two to six qubits of the Heisenberg model with $h = 0.5$ and $\Delta = -0.5, 0.0, 0.5$. A total of 100 runs are made for each point, with the optimal state taken to be the one that maximizes the fidelity.}
    \label{fig:heisenberg_shots}
\end{figure*}

The next step was to carry out shot-based simulations for the Heisenberg model with $h = 0.5$ and $\Delta = -0.5, 0.0, 0.5$, respectively, as shown in Fig.~\ref{fig:heisenberg_shots}. Using \ac{SPSA}, ten runs were carried out for each $\beta$, while the number of iterations was taken to be $100n$ for each run, with $2n$ function evaluations at each iteration to estimate the gradient in $n$ random directions, i.e. $200n^2$. Similar to the number of layers, the choice of the number of function evaluations was heuristically chosen so that the scaling is polynomial. The number of commuting sets of Pauli strings of the Heisenberg model is three. Furthermore, each circuit was also measured with 1024 shots. and the \texttt{M3} package~\cite{Nation2021} was also utilized to perform error mitigation. Table~\ref{tab:scaling_6} shows a summary of the optimization scaling.

\begin{table}[t]
\centering
\resizebox{\linewidth}{!}{
\begin{tabular}{|l|l|l|}
    \hline
    \# of iterations for each run & $100n$ & $\cO(n)$ \\
    \hline
    \# of function evaluations for each run & $200n^2$ & $\cO(n^2)$ \\
    \hline
    \# of circuits per function evaluation & $3$ & $\cO(1)$ \\
    \hline
    \# of circuit evaluations for each run & $600n^2$ & $\cO(n^2)$ \\
    \hline
    \# of shots for each circuit evaluation & $1024$ & $\cO(1)$ \\
    \hline
\end{tabular}
}
\caption{Scaling of \ac{SPSA} for shot-based simulations.}
\label{tab:scaling_6}
\end{table}

Naturally, the choice of optimizer, along with the finite number of measurements used to reconstruct both the von Neumann entropy and the expectation value, are shown to affect the performance of the \ac{VQA}. Nevertheless, while the results of Fig.~\ref{fig:heisenberg_shots} exhibit a lower, albeit relatively high fidelity $F \gtrsim 0.93$, the \ac{VQA} shows notable promise in being able to produce Gibbs states of complex interacting Hamiltonians, such as the Heisenberg \textit{XXZ} model.

\section{Conclusion} \label{sec:conclusion}

We addressed the preparation of a thermal equilibrium state of a quantum many-body system on a \ac{NISQ} device. We exploited the uniqueness of the Gibbs state as the state that minimizes the Helmholtz free energy, thus providing a faithful objective function for a \ac{VQA}.

The novelty of the proposed \ac{VQA} consisted in splitting the \ac{PQC} in two parameterized unitaries, one acting on an ancillary register, and one on a system register. The former is tasked with determining the Boltzmann weights of the Gibbs distribution, corresponding to a given temperature, while the latter performs the rotation from the computational basis to the energy basis of a given Hamiltonian.

We benchmarked our \ac{VQA} preparing the Gibbs state of the transverse field Ising model and obtained fidelities $F \simeq 1$ for system sizes up to six qubits in statevector simulations, across a broad range of temperatures, with a slight dip at intermediate ones. Moreover, we tested our \ac{VQA} on the Heisenberg model with a transverse field, similarly obtaining fidelities $F \simeq 1$ in statevector simulations. However, performance on current \ac{NISQ} devices, investigated both by noisy simulations and real-hardware execution on IBM devices, showed a degradation in the results of the \ac{VQA} with increasing system size. This may have been caused by the limited connectivity and the noise present in the device. Nevertheless, executing our \ac{VQA} on \ac{NISQ} devices still provides an improvement upon the recent developments in variational Gibbs state preparation, see e.g., Ref.~\cite{Sagastizabal2021}.

It is important to notice that the structure of our \ac{VQA} does not depend on the specific Hamiltonian to be tackled, nor on any prior knowledge of its spectrum. For example, the structure of the unitary $U_S(\bm{\varphi})$ could be adjusted in order to match some specific features of the eigenstates (if these are known), or the parameterized unitary $U_A(\bm{\theta})$ could be replaced by a deterministic procedure (e.g., the one reported in \cite{Sannia2023}), if the probabilities of the Boltzmann distribution are known. 

However, even without requiring any such knowledge, our `Hamiltonian-agnostic' variational approach gives an effective way to prepare Gibbs states of arbitrary quantum many-body systems on a quantum computer, providing an advancement over previous methods, especially thanks to the modular structure of our \ac{PQC}. This could significantly contribute to both performing quantum thermodynamical experiments on a quantum computer, as well as faithfully preparing Gibbs states to be used in a great variety of computational tasks. Furthermore, preparing moderately-sized many-body systems with our \ac{VQA} may also be sufficient for exploring finite-sized effects of certain physical models~\cite{Um2007}.

Some final remarks: many modular elements of the \ac{VQA} have the capacity to be significantly improved. While the scope of this manuscript was to provide a proof-of-concept \ac{VQA} for preparing Gibbs states by estimating the entropy directly without any truncation, we will mention potential avenues for future research. In particular, more robust error mitigation techniques could be implemented, such as those present in the \texttt{mitiq} library~\cite{mitiq}. In addition to this, we delve into the consequences of working with a limited number of samples when attempting to estimate the entropy in Appendix~\ref{sec:entropy_estimation}. Moreover, we explore an alternative entropy estimation technique that exhibits promise in its potential to scale beyond the \ac{NISQ} era of computing. Barren plateaus in deep \acp{PQC} and noisy devices are also a considerable challenge to address. In Appendix~\ref{sec:barren_plateau}, we qualitatively discussed the requirements needed to investigate barren plateaus for our algorithm, as well as the \acp{PQC} that stem from its particular structure. Lastly, it is worth noting that the choice of optimizer plays a pivotal role in the performance of the \ac{VQA}, particularly in the presence of noise. An in-depth analysis of various optimizers could significantly enhance the reliability of the \ac{VQA}.

The Python code for running the statevector simulations, using Qulacs~\cite{Suzuki2021}, and the noisy simulations, as well as the Runtime Program, using Qiskit~\cite{qiskit}, can be found on GitHub~\cite{github}.

\section*{Acknowledgments}

M.C. acknowledges fruitful discussions with Jake Xuereb and Felix Binder on the thermodynamical concepts of the manuscript. M.C. and T.J.G.A. would like to thank Matteo Rossi, Marco Cattaneo and Zoe Holmes for the interesting discussions on the algorithmic component of the Manuscript. M.C. would also like to thank Fiona Sammut for discussions on estimating the entropy. M.C. acknowledges funding by TESS (Tertiary Education Scholarships Scheme), and project QVAQT (Quantum variational algorithms for quantum technologies) REP-2022-003 financed by the Malta Council for Science \& Technology, for and on behalf of the Foundation for Science and Technology, through the FUSION: R\&I Research Excellence Programme. T.J.G.A. acknowledges funding from the European Commission via the Horizon Europe project ASPECTS (Grant Agreement No. 101080167). The views and opinions expressed are however those of the author(s) only and do not necessarily reflect those of the European Union. Neither the European Union nor the granting authority can be held responsible for them. J.G. is funded by a Science Foundation Ireland-Royal Society University Research Fellowship and his work is also supported by the European Research Council Starting Grant ODYSSEY (Grant Agreement No. 758403). S.L. acknowledges support by MUR under PRIN Project No. 2022FEXLYB Quantum Reservoir Computing (QuReCo). C.M. acknowledges funding by the European Union PNRR National Centre on HPC, Big Data and Quantum Computing, PUN: B93C22000620006. We acknowledge the use of IBM Quantum services for this work; the views expressed are those of the authors, and do not reflect the official policy or position of IBM or the IBM Quantum team.

\appendix

\section{Error Analysis of Entropy Estimation} \label{sec:entropy_estimation}

In general, reconstructing the probability distribution faithfully, requires an exponential number of shots, and particularly free fermion distributions can be hard to learn~\cite{Nietner2023}. However, let us look at estimating the entropy using the \ac{ML} estimator, rather than focusing on the reconstruction of the probability distribution. The \ac{ML} estimator was shown to have a bias and variance that in general decreases as $\cO(N^{-1})$ for $N \gg M$~\cite{Paninski2003}. The outcome of one shot of a quantum circuit can be described by a multinomial distribution $\vec{p}$, where $p_i$ is the probability of observing a bit string $i$. Given $N$ shots, filling $M$ bins, the \ac{ML} estimator~\cite{Roulston1999, Paninski2003} of the von Neumann entropy is given by
\begin{equation}
    \cS_\text{ML} = -\sum_{i=1}^M q_i \log q_i,
\end{equation}
were $q_i = n_i / N$, such that $n_i$ represents the number of times the bit string $i$ appears after $N$ shots. The variance of the entropy can be easily computed as
\begin{equation}
    \mathbb{V}(\cS_\text{ML}) = \sum_i^M (1 + \log q_i)^2 \mathbb{V}(q_i)
    \label{eq:errEntropy}
\end{equation}
where $\mathbb{V}(q_i) \approx q_i (1 - q_i) / N$. In fact, it can be shown that for all $N$, and all possible distributions, the variance of the \ac{ML} estimator for entropy is bounded above as
\begin{equation}
    \mathbb{V}(\cS_\text{ML}) \leq \frac{(\log(N))^2}{N},
\end{equation}
as proven in Ref.~\cite{Antos2001}, and that
\begin{equation}
    P(|\cS_\text{ML} - \mathbb{E}(\cS_\text{ML})| > \epsilon) \leq 2e^{-\frac{N \epsilon^2}{2(\log(N))^2}}.
\end{equation}
It is important to note that this bound is not particularly tight, and it is independent of $M$ and the probability distribution. Moreover, the \ac{ML} estimator was proven to be negatively biased everywhere, and that
\begin{equation}
    \mathbb{E}_{\vec{p}}(\cS_\text{ML}) \leq \cS(\vec{p}),
\end{equation}
where $\mathbb{E}_{\vec{p}}$ denotes the conditional expectation given $\vec{p}$, and that equality is only achieved when $\cS(\vec{p}) = 0$, meaning the distribution is supported on a single point (or in the case of Boltzmann distributions for $\beta \rightarrow \infty$). In the case of $N \gg M$, the Miller-Madow bias correction~\cite{Roulston1999, Paninski2003} gives that
\begin{equation}
    \cS(\vec{p}) = \cS_\text{ML}(\vec{p}) + \frac{M - 1}{2N} + \cO(N^{-1}).
\end{equation}

In the case of \ac{NISQ} and future quantum algorithms, given that the Hilbert space of qubits grows as $M = 2^n$, we can reasonably assume that $N \ll M$ as soon as $n > 20$. As a result, we need to look towards entropy estimation techniques when we are in a heavily undersampled regime. Bayesian inference is a typically employed method in these situations. While learning a probability distribution might generally require an exponential number of samples~\cite{Nietner2023}, computing functionals of such distributions might not. As a result, the \ac{NSB} estimator employs Bayesian inference to obtain both the entropy and its a posteriori standard deviation.  We utilize the Python package \texttt{ndd}~\cite{ndd} to compute the $\cS_\text{NSB}$, while referring interested readers to Refs.~\cite{Nemenman2002, Nemenman2011} for the details.

\begin{figure*}[t]
    \centering
    \includegraphics[width=\textwidth]{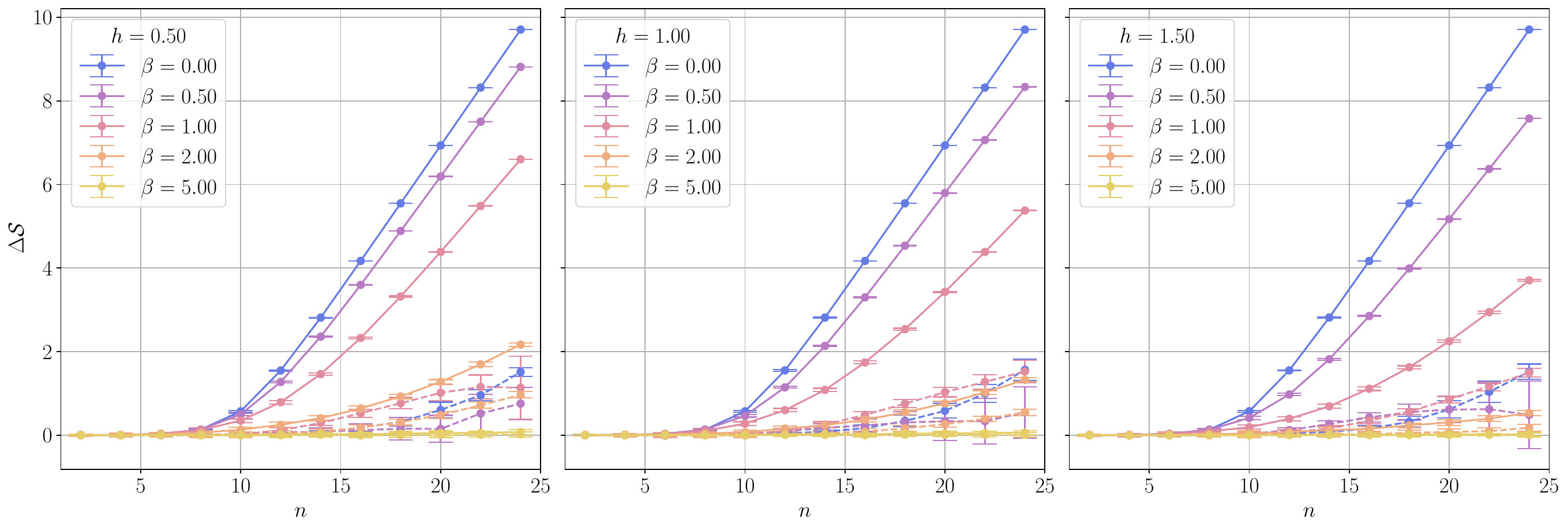}
    \caption{Average additive error $\Delta\cS$ (bias) in entropy estimation as a function of the number of qubits $n$, using the \ac{ML} (solid lines) and \ac{NSB} (dashed lines) estimators, where the error bars represent the standard deviation. The number of shots is $N = 1024$ and the number of samples for each point is $100$, with $h = 0.5, 1.0, 1.5$ for the Ising model, respectively.}
    \label{fig:entropy}
\end{figure*}

Fig.~\ref{fig:entropy} shows the results of using the \ac{ML} and \ac{NSB} estimators using a finite amount of shots $N = 1024$, for the Ising model with $h = 0.5, 1.0, 1.5$, respectively. In particular we show how the additive error (bias) of the entropy scales as the number of qubits increases. Each point in Fig.~\ref{fig:entropy} is obtained by averaging the entropy over $100$ samples, with the error bars representing the standard deviation.

As one can expect, for a number of qubits $n$ such that $2^n \ll N$, both \ac{ML} and \ac{NSB} estimators obtain a bias and standard deviation close to zero, for all values of $\beta$. There is a transition region where $2^n \sim N$, where the bias, particularly for low values of $\beta$, starts to increase. In the region where $2^n \gg N$, the \ac{ML} estimator is only valid in the large $\beta$ regime, since there is usually a finite number of non-zero probabilities which is much smaller than $N$. It is important to note that although the error is increasing linearly with the number of qubits, the number of shots $N$ needed to reduce the bias to a constant error increases exponentially as a function of the number of qubits $n$. Specifically, in the region $2^n \gg N$, the \ac{ML} estimator reaches the upper bound of $\log(N)$, and so for $\beta = 0$, $\Delta\cS_\text{ML} = \log(2^n) - \log(N) = \log(2^n/N)$.

On the other hand, the \ac{NSB} estimator obtains a much lower bias, at the cost of a slightly higher standard deviation. While the behavior of the \ac{NSB} estimator is hard to surmise given that exact diagonalization results allowed us to see until $n = 24$, there are a few instances where for intermediate and high $\beta$ the estimator seems to flatten, or even decrease as $n$ increases. Proving that for a particular $\beta$, the \ac{NSB} estimator reliably acquires a bias that scales as $\cO(\text{poly}(\log(n)))$, would mean that using a number of shots $N$ that scales polynomially would achieve a constant additive error, implying scalability in entropy estimation.

\section{Necessity of Entangling Gates in \texorpdfstring{$U_A$}{the Ancilla Unitary}} \label{sec:ancilla_ansatz}

In the main text, we specified that we required entanglement in the ancillary register to be able to prepare the Boltzmann distribution of the Ising model. While we only used one layer of a hardware-efficient ansatz, we concluded that least one entangling layer is necessary for preparing the Boltzmann distribution of the Ising model, and we will show this by considering the converse. Suppose the ancilla ansatz is only composed of local $R_y$ gates, then we get
\begin{align}
    \bigotimes_{i=0}^{n-1}R_y(\theta_i)\ket{0}_i &= \bigotimes_{i=0}^{n-1} \left( \cos\left(\frac{\theta_i}{2}\right)\ket{0}_i + \sin\left(\frac{\theta_i}{2}\right)\ket{1}_i\right) \nonumber \\
    &= \sum_{i=0}^{d-1} \prod_{j \in S_{i=0}} \cos\left(\frac{\theta_j}{2}\right) \prod_{k \in S_{i=1}} \sin\left(\frac{\theta_k}{2}\right) \ket{i} \nonumber \\
    &= \sum_{i=0}^{d-1} p_i \ket{i},
\end{align}
where $S_{i=0} \equiv \left\{j~|~j = 0~\forall~\textrm{bits}~j \in i \right\}$, that is, the set of bits in $i$ which are equal to $0$, and similarly defined for $S_{i=1}$, and $\ket{i}$ is the computational basis state. This implies that
\begin{equation}
    \prod_{j \in S_{i=0}} \cos\left(\frac{\theta_j}{2}\right) \prod_{k \in S_{i=1}} \sin\left(\frac{\theta_k}{2}\right) = p_i.
\end{equation}
Now, without loss of generality, consider these specific cases:
\begin{subequations}
\begin{align}
    &\prod_{j=0}^{n-1} \cos\left(\frac{\theta_j}{2}\right) = p_0, \\
    &\prod_{j=0}^{n-2} \cos\left(\frac{\theta_j}{2}\right) \sin\left(\frac{\theta_{n-1}}{2}\right) = p_1, \\
    &\prod_{j=0}^{n-3} \cos\left(\frac{\theta_j}{2}\right) \cos\left(\frac{\theta_{n-1}}{2}\right) \sin\left(\frac{\theta_{n-2}}{2}\right)  = p_2, \\
    &\prod_{j=0}^{n-3} \cos\left(\frac{\theta_j}{2}\right) \sin\left(\frac{\theta_{n-2}}{2}\right) \sin\left(\frac{\theta_{n-1}}{2}\right) = p_3.
\end{align}
\end{subequations}
Combining the above equations results in
\begin{subequations}
\begin{align}
     \frac{p_0}{p_1} &= \frac{\cos\left(\frac{\theta_{n-1}}{2}\right)}{\sin\left(\frac{\theta_{n-1}}{2}\right)}, \\
    \frac{p_0}{p_2} &= \frac{\cos\left(\frac{\theta_{n-2}}{2}\right)}{\sin\left(\frac{\theta_{n-2}}{2}\right)}, \\
    \frac{p_0}{p_3} &= \frac{\cos\left(\frac{\theta_{n-2}}{2}\right) \cos\left(\frac{\theta_{n-1}}{2}\right)}{\sin\left(\frac{\theta_{n-2}}{2}\right) \sin\left(\frac{\theta_{n-1}}{2}\right)},
\end{align}
\end{subequations}
and finally, combining the above equations implies that
\begin{align}
    \frac{p_0}{p_1}\frac{p_0}{p_2} = \frac{p_0}{p_3} &\implies \frac{p_0}{p_1 p_2} = \frac{1}{p_3} \nonumber \\
    &\implies p_0 p_3 = p_1 p_2 \nonumber \\
    &\implies e^{-\beta E_0} e^{-\beta E_3} = e^{-\beta E_1} e^{-\beta E_2}.
\end{align}
Applying logs to both sides and simplifying, we get
\begin{equation}
    E_0 + E_3 = E_1 + E_2,
    \label{eq:constraint}
\end{equation}
which is not in general true for the Ising model. The above reasoning can be adjusted to obtain further constraints in the manner of Eq.~\eqref{eq:constraint}.

\section{Barren Plateau Analysis} \label{sec:barren_plateau}

By following the analysis carried out in Ref.~\cite{Cerezo2021}, we qualitatively discuss the trainability of our \ac{VQA}. We can decompose our cost function as
\begin{widetext}
\begin{equation}
    \cF(\rho_S) = \Tr{\cH \rho_S} - \beta^{-1} \cS(\rho_A) = \Tr{\cH U_S \rho_S' U_S^\dagger} + \beta^{-1} \sum_{i=0}^{d-1} \Tr{O_i U_A \ketbra{0}_{A}^{\otimes n} U_A^\dagger} \ln \Tr{O_i U_A \ketbra{0}_{A}^{\otimes n} U_A^\dagger},
    \label{eq:cost_function}
\end{equation}
\end{widetext}
where $\rho_S' = \PTr{A}{V \ketbra{0}_{AS}^{\otimes 2n} V^\dagger}$, $V = \textsc{CNOT}_{AS}(U_A \otimes \dI_S)$ and $O_i = \ketbra{i}$. Now, Ref.~\cite{Cerezo2021} only considers cost functions of the form
\begin{equation}
    C = \Tr{O U \rho U^\dagger},
    \label{eq:cerezo_cost_function}
\end{equation}
where $\rho$ is an arbitrary quantum state of $n$ qubits, $O$ is any operator, and $U$ is an alternating layered ansatz. Given that our cost function in Eq.~\eqref{eq:cost_function} is not in the form of Eq.~\eqref{eq:cerezo_cost_function}, because of the logarithm in the von Neumann entropy, our comparison should be taken solely as a qualitative discussion on the possibility of barren plateaus.

Now with reference to Eq.~\eqref{eq:cost_function}, we have that $\cH$ is 2-local, while $O_i$ is 1-local, and both $U_A$ and $U_S$ are alternating layered ans\"atze. Theorem 2 of Ref.~\cite{Cerezo2021} gives a lower bound on the variance of the gradient of the cost function as a function of the number of layers, and as such, the trainability of the \ac{PQC}. If the number of layers $l = \cO(\log(n))$, then the variance vanishes no faster than polynomially, hence making the \ac{PQC} trainable. If the number of layers $l = \cO(\text{poly}(\log(n)))$, then the variance vanishes faster than polynomially, but no faster than exponentially, settling in a transition region in between trainable and untrainable.

In the case of $\beta \rightarrow \infty$, our cost function equates directly to Eq.~\eqref{eq:cerezo_cost_function} (since the \ac{VQA} effectively reduces to a \ac{VQE}, and thus we require $l = \cO(\log(n))$ for our circuit to be trainable. On the other hand, in the case of $\beta \rightarrow 0$, the cost function simplifies to maximizing the von Neumann entropy, which is acquiring the maximally mixed state. While we cannot directly relate the von Neumann entropy as a cost function with Eq.~\eqref{eq:cerezo_cost_function}, we have numerically seen that preparing the mixed state is a relatively straightforward task. Nevertheless, analysis on the trainability of utilizing the von Neumann entropy as the (or part of the) cost function should be sought to be able to detect the presence of barren plateaus.

Now for any finite $\beta > 0$, the problem of determining whether a barren plateau is possible for the generalized free energy is out of the scope of this work. Nevertheless, we can possibly surmise that, given $U_A$ and $U_S$ being alternating layered ans\"atze, with $\cH$ being a 2-local Hamiltonian consisting of traceless operators, then using a number of layers for both $U_A$ and $U_S$ that scales at most as $l = \cO(\log(n))$, might result in a \ac{PQC} that is trainable. This would hold if Theorem 2 of Ref.~\cite{Cerezo2021} also holds for cost functions in the form of Eq.~\eqref{eq:cost_function}.

\bibliographystyle{apsrev4-2}
\bibliography{ref.bib}

\end{document}